# AN IMPROVED GENERIC ER SCHEMA FOR CONCEPTUAL MODELLING OF INFORMATION SYSTEMS

Pieris D[1], Wijegunasekera M.C[2], Dias N. G. J[3]

[1]Department of HRM, Faculty of Commerce and Management Studies [2]Department of Software Engineering, Faculty of Computing and Technology [3]Department of Computer Systems Engineering, Faculty of Computing and Technology
University of Kelaniya, Sri Lanka

[1]mdp@kln.ac.lk, [2]carmel@kln.ac.lk, [3]ngjdias@kln.ac.lk

## 1. Abstract

The Entity Relationship (ER) model is widely used for creating ER schemas for modelling application domains in the field of Information Systems development. However, when an ER schema is transformed to a Relational Database Schema (RDS), some important information on the ER schema may not be represented meaningfully on the RDS. This causes loss of information during the transformation process. Although, several previous researches have proposed solutions to remedy the situation, the problem still exists. Thus, in this on-going research we wish to improve the proposed solutions and maximize information preservation in the ER to relational transformation process. Cardinality ratio constraints, role names, composite attributes, and certain relationship types are among the information frequently lost in the transformation process. Deficiencies in the ER model and the transformation method seems causing this situation. We take the view that if the information lost is resolved; a one-to-one mapping should exist from the ER schema to its RDS. We modified the ER model and the transformation algorithm following a heuristic research method with a view to eliminating the deficiencies and thereby achieving a one-to-one mapping. We should show that the mapping exists for any real world application.

We create a generic ER schema - an ER schema that represents any phenomena in symbolic form - and use it to show that a one-to-one mapping exists for any real world application. In this paper, we explore our generic ER schema and its advantages over its predecessors in view of representing any real world application.

## 1. Introduction

When an Information Systems development work is undertaken a conceptual model is drawn in the form of an ER schema using the ER model[1, 2] to represent user requirements of the application domain concerned. An ER schema is a graphical diagram and represents phenomena in the real world, such as entities, relationships, and attributes via graphical constructs, for instance, rectangles, diamonds, and ovals, etc. The constructs that are modelled on an ER schema are usually named by the corresponding real-world names (e.g., Employee, Designation, and Location, as in Figure 1) occurring in the application domain to which the ER schema is drawn.

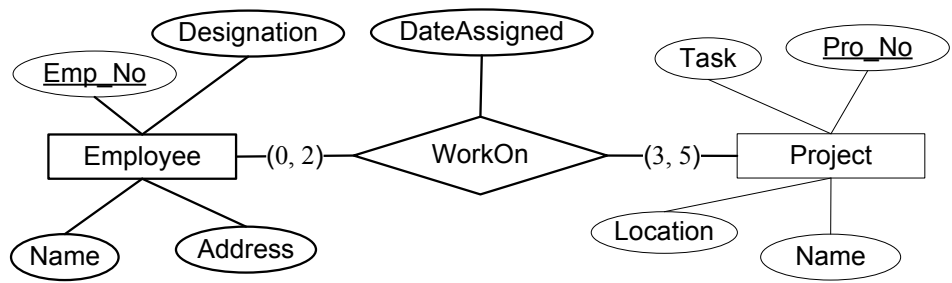

Figure 1: An ER schema that represents a real-world application domain

The ER notation using real-world names shown in Figure 1 (adapted from Elmasri & Navathe [3]) is known to be popular, natural, and understandable. Many other authors [4-6] often use the same or a similar version of it for ER schema modeling.

Even though the ER notations using real-world names are popular for representing real-world situations, they also pose limitations when they are used for automation of the ER to relational model[7, 8] transformation process. After an ER schema is created it is then transformed to the relational database schema (RDS). However, an information loss is occurred during this transformation process [9-11]. This information loss is difficult to understand and resolve unambiguously using real-world ER schemas. The information loss identified and the solution provided with regard to an ER schema representing one application domain (e.g. Company scenario, Figure 1) may not be relevant with regard to another ER schema representing a different application domain

(e.g., Library database). Therefore, an ER schema expressed in a formal notation and that can represent any application domain, in general, is necessary for addressing the information loss. In other words, an ER schema independent of any real-world application domain is required.

Several researchers have proposed some generic ER schemas to address similar issues (e.g., [12], [13]). The proposed generic ER schemas never use real-world names, but instead alphanumeric symbols for naming their constructs. Nevertheless, proposed generic ER schemas also commit limitations. In the current study, we explore a generic ER schema that we have undertaken to develop for use in our main research.

Accordingly, in section 2, we discuss the generic ER schemas proposed by two researchers and the limitations of them. Section 3 presents the preliminaries of our method for developing a generic ER schema. Section 4 describes how relationship types and attributes are represented in the generic ER schema. Section 5 deals with how the structural constraints that are associated with relationship types can be represented. Finally, Section 6 presents the conclusion and future research that we plan to undertake using our generic ER schema that we will unfold.

## 2. Generic ER schemas proposed by some researchers

Storey [12] proposed a model for a generic ER schema, as follows:

*"... Let $E = \{E_i\}$ be the set of all entity types and $A = \{A_{ij}\}$ the set of all attributes where $A_{ij}$ is the $j^{th}$ attribute of the $i^{th}$ entity type." (p. 4)*

This model specifies how several regular entity types and attributes attached to them can be represented. However, the author has not given any formal way of representing Primary Key(PK) attributes and relationship types in her work.

Atzeni, et al [13] shown that their proposed generic ER schema as given in Figure 2. This schema does not indicate how several relationship types can be represented between entity types and how they can be named, uniquely and consistently. For example, assume that a second relationship type exists between $E_0$ and $E_1$ and is named as $R_2$. Further, assume that another third relationship type exists between either $E_0$ and $E_1$ or $E_0$ and another entity type $E_2$. Then, it is not clear how this third relationship type can be named.

The ER schema indicates PK attributes and non-PK simple attributes using two separate visual constructs. For instance, $A_{01}$ of $E_0$, and $A_{11}$ of $E_1$ are closed circles, and they indicate PK attributes of the respective entity types. The remaining simple attributes are represented as open circles. A PK given in a visual notation may not be able to understand if it is mentioned outside the ER schema. When the PK attributes are listed as: $A_{01}$, $A_{11}$ the visual method is absent. However, still, they could sometimes be identified as PKs because of their second prefix being 1, always. If this is the case, it indicates a method redundancy. On the other hand, the numbers appear to have been assigned with multiple tasks: one is counting, and the other one is representing a semantic meaning - being a PK.

Further, the schema has not provided a method for representing attributes attached to relationship types.

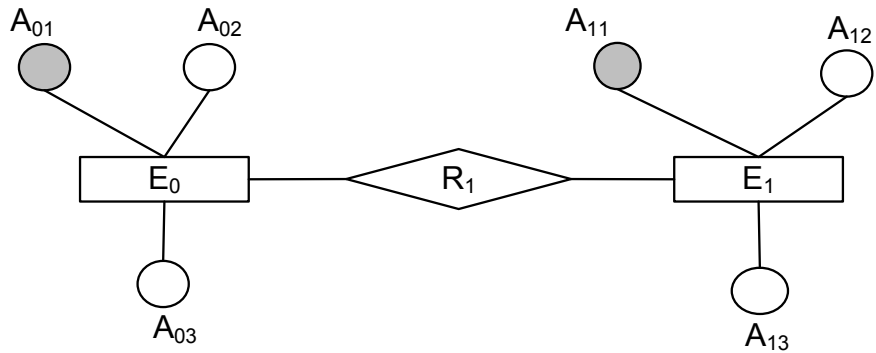

Figure 2: A generic ER schema that represents real-world phenomena in alphanumeric (symbolic) form

Source: Adapted from Atzeni, Ceri, Paraboschi, & Torlone [13]

In the current research, we propose a method for developing a generic ER schema that can overcome the above-mentioned issues.

## 3. Prefaces of a method for developing a Generic ER schema

In our proposed model, the letter "$e$" represents a regular entity type. Accordingly, $e_i$ represents the $i^{th}$ regular entity type, where $i \in \mathbb{N}$ – Natural number set, i.e., $\mathbb{N} \equiv \{1, 2, 3 \ldots\}$. Consider an entity type and an attribute belonging to it. An attribute belonging to its entity type is an association that exists between the attribute and its entity type. Therefore, we consider the attribute as a mapping $s$ from the entity type. Then the simple attribute of the entity type $e_i$ can be denoted as $s(e_i)$. In general, we denote the mapping as $s_j(e_i)$ such that $s_j$ represents the $j^{th}$ simple attribute of the $i^{th}$ entity type, where $i, j \in \mathbb{N}$. Accordingly, the simple attributes of $e_1$ can be denoted as $s_1(e_1)$, $s_2(e_1)$, $s_3(e_1)$, ... , $s_n(e_1)$. Here, $s_1, s_2, s_3,$ ...$s_n$ are separate mappings. Each separate mapping $s_i$, is defined from the common set $\{e_1, e_2, e_3, ..., e_n\}$ of all the regular entity types of the ER schema to the common set $\{s_1(e_1), s_2(e_1),..., s_1(e_2), s_2(e_2),,..., s_1(e_3), s_2(e_3), ..., s_1(e_n), s_2(e_n)\}$ of all the simple attributes of it.

For example, consider the Employee entity type, and its simple attributes, Name, Address, and Designation, as given in Figure 1. According to our mapping the attributes are stated as $s_1$ (Employee) = Name, $s_2$ (Employee) = Address, and $s_3$ (Employee) = Designation.

In this research, we limit the PK to be a single attribute and not a composite attribute. We define a rule for setting the name of an entity type's PK attribute. The PK attribute's name should be related to the entity type's name. Accordingly, the PK name should be the name of the entity type or its first three letters or its first three letters with any identifiable string of the entity name concatenated with the underscore and any other suitable identifiable string. For example, the PK's name of the entity type Employee (Figure 1) can be set as either Employee_No or Emp_No or Empye_No, etc. In conclusion, whatever a name is chosen for a PK it should be related to the name of the entity type to which it belongs to and unique to the entire ER schema.

Consequently, we consider the PK of an entity type to be a mapping of it. Let this mapping to be $k$, and declare the PK of the entity type $e_i$ as $k(e_i)$. For instance, if assume that the PK of the entity type Employee (Figure 1) is to be Emp_No. Then, $k$ (Employee) = Emp_No. Figure 3, below, shows a generic ER schema that follows the rules described above. The schema contains a single regular entity type, $e_i$, and a set of $n$ ($n \in \mathbb{N}$) number of simple attributes.

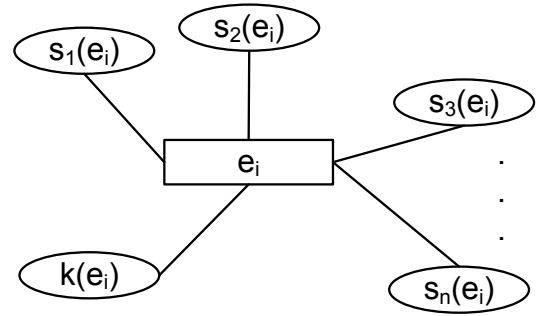

Figure 3: A generic ER schema created under the current research

## 4. Representation of relationship types and attributes attached to them

This section describes a method for representing binary relationship types existing between regular entity types and attributes attached to them. Let $r$ be a binary relationship type that exists between two regular entity types: $e_i$ and $e_t$. We denote this relationship type as $r(e_i.e_t)$. If more relationship types exist between $e_i$ and $e_t$, they can be denoted as $r_v(e_i.e_t)$, where $v \in \mathbb{N}$. Accordingly, several relationship types, if they exist, between $e_i$ and $e_t$ can be represented as $r_1(e_i.e_t), r_2(e_i.e_t), \ldots$, and so on. An attribute attached to the relationship type, $r_v(e_i.e_t)$ can be denoted as $s_u(r_v(e_i.e_t))$, where $u, v, i, t \in \mathbb{N}$.

For example, a binary relationship type that exists between the entity types, $e_1$ and $e_2$ can be denoted as $r_1(e_1.e_2)$ while an attribute attached to the relationship type, $r_1(e_1.e_2)$ can be denoted as $s_1(r_1(e_1.e_2))$. A second attribute, if exists to the relationship type, can be denoted as $s_2(r_1(e_1.e_2))$. The following Figure 4 shows these developments.

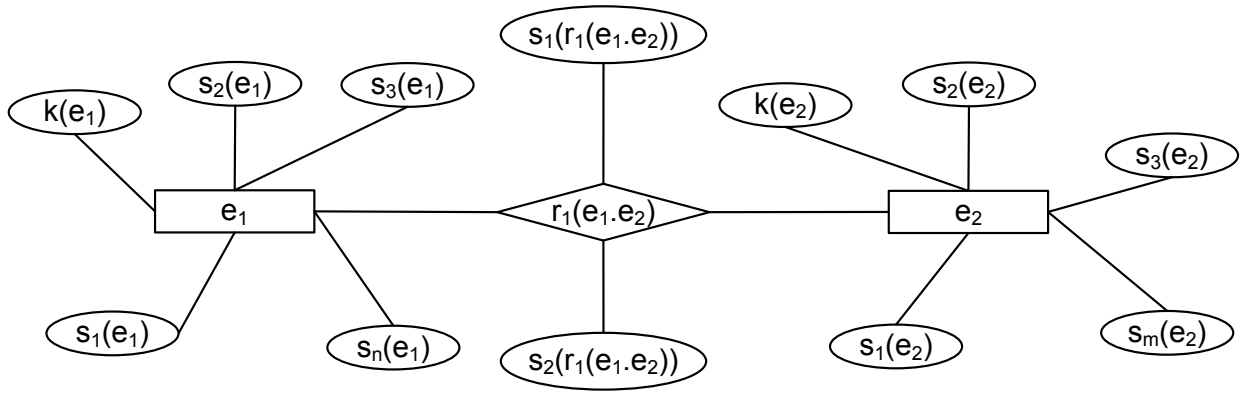

Figure 4: How a relationship type that exists between two regular entity types can be represented together with its attributes attached, in the proposed generic ER schema

## 5. Representation of structural constraints

The ER schema (Figure 1) shows structural constraints, such as (0, 2) and (3, 5) indicated at both sides of the relationship type, WorksOn.

Accordingly, two pairs of $(min, \max)$ structural constraints[3] often exist at both sides of the relationship type $r_1(e_1.e_2)$ in the

ER schema, Figure 4. One of the two pairs exists at the side $e_1$ of the relationship type $r_1(e_1.e_2)$, while the other pair exists at the side $e_2$ of it. We denote the side of $e_1$ of the relationship type $r_1(e_1.e_2)$ as $e_1.r_1(e_1.e_2)$, uniquely and consistently. Similarly, the entity type $e_2$'s side can be denoted as $e_2.r_1(e_1.e_2)$. Consequently, the structural constraint either $min$ or $max$ of each side can be defined as a mapping of that side. Let the $min$ constraint be denoted by the letter $m$. Then the $min$ constraint that exists at the $e_1.r_1(e_1.e_2)$ side of the relationship type can be defined as $m(e_1.r_1(e_1.e_2))$. Similarly, given that $max$ constraint is denoted as $x$, then the $max$ constraint at the same side can be denoted as $x(e_1.r_1(e_1.e_2))$. In the same way the $min-max$ constraints that exist at $e_2.r_1(e_1.e_2)$ side can be denoted as $m(e_2.r_1(e_1.e_2))$ and $x(e_2.r_1(e_1.e_2))$. The Figure 5 shows how these structural constraints can be represented on the ER schema.

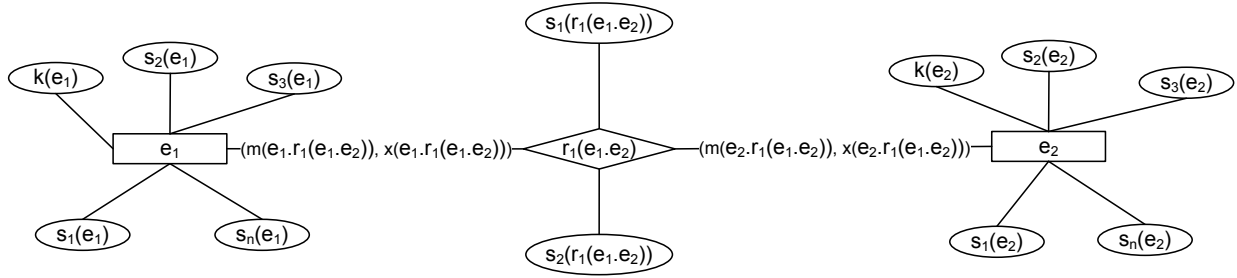

Figure 5: Representation of structural constraints in a proposed generic ER schema

## 6. Conclusion and future work

The generic ER schema proposed above has the ability to overcome most of the limitations appearing to exist in its predecessor ones. The proposed generic schema can represent regular entity types, relationship types and attributes attached to them, and structural constraints, uniquely and consistently. The uniqueness and consistency of them are retained even if they are represented only by their symbolic labels and outside the ER schema. For instance, consider the list of labels obtained from the ER schema in Figure 5, as follows: $e_2$, $r_1(e_1.e_2)$, $s_2(e_2)$, $x(e_2.r_1(e_1.e_2))$, $k(e_1)$, $s_2(r_1(e_1.e_2))$. We believe that a person with a clear understanding of the logic (sections: 3, 4, and 5) used to create the generic ER schema should be able to identify and determine the labels' corresponding ER constructs, uniquely and consistently.

In future research, we will focus on how our generic ER schema can be transformed to an RDS. Subsequently, we will obtain a one-to-one mapping from the generic ER schema to the RDS. We will show that the information is retained and its loss is resolved and the proposed approach is valid for any real-world application domain.